\documentstyle[11pt,epsfig,psfig]{article}

\makeatletter
\@addtoreset{equation}{section}

\makeatother
\def\beq{\begin{equation}}
\def\eeq{\end{equation}}

\def\ds{\displaystyle}
\def\text#1{\mbox{\scriptsize #1}}

\begin{document}

\bigskip \noindent {\Large {\bf
Vacuum densities and zero-point energy for fields obeying Robin
conditions on cylindrical surfaces }}

\vspace*{5mm} \noindent August Romeo$^{a,}$\footnote{%
E-mail: romeo@ieec.fcr.es} and Aram A. Saharian$^{b,}$\footnote{%
E-mail: saharyan@server.physdep.r.am} \newline
\noindent ${}^a$ Institut d'Estudis Espacials de Catalunya (IEEC/CSIC),
Institut de Ci{\`e}ncies de l'Espai (CSIC), \newline
Edifici Nexus-201 - c. Gran Capit\`a 2-4, 08034 Barcelona \newline
\noindent ${}^b$ Department of Physics, Yerevan State University, 1 Alex
Manoogian St, 375049 Yerevan, Armenia \vspace*{10mm}

\noindent {\bf Abstract.} The Casimir effect for general Robin
conditions on the surface of a cylinder in $D$-spacetime
dimensions is studied for massive scalar field with general
curvature coupling. The energy distribution and vacuum stress are
investigated. We separate volumic and superficial energy
contributions, for both interior and exterior space regions. The
possibility that some special conditions may be energetically
singled out is indicated.

\section{Introduction}

The whole scientific production related to the Casimir effect has already a
quite impressive volume (for reviews see e.g. refs.\cite{PMG}). This effect
describes the energy variation undergone by a quantum system in which the
field modes are constrained by specific conditions on a given boundary. The
most often studied types of boundary and conditions are those
associated to well known problems, e.g., plates, spheres, and vanishing
conditions, perfectly conducting conditions, etc.

Our aim in the present paper is to analyze a combination of
boundary geometry and conditions which hasn't attracted the
researcher's attention up to now. We are referring to general
homogeneous ---or 'Robin'--- boundary conditions on infinitely
long cylindrical surfaces. Such conditions are an extension of
the ones imposed on perfectly conducting boundaries and may, in
some geometries, be useful for depicting the presence of
interfaces between dielectric media. Robin type of boundary
conditions also appear in considerations of the vacuum effects
for a confined charged scalar field in external fields
\cite{Ambjorn2} and in quantum gravity \cite{Moss,Espos}. Unlike
'purely-Neumann' boundary conditions, they can be made
conformally invariant. The Casimir energy for cylindrical
geometry can be important to the flux tube model of confinement
\cite{Fishbane,Barbash} and for determining the structure of the
vacuum state in interacting field theories \cite{Ambjorn1}. The
cylindrical problem with perfectly conducting conditions was
first considered in ref.\cite{DRM} (see also refs.\cite{GR},
\cite{MNN}). While the earliest studies have focused on global
quantities, such as the total energy and stress on a shell, the
local characteristics of the corresponding electromagnetic vacuum
are considered in \cite{Sah1cyl} for the interior and exterior
regions of a conducting cylindrical shell, and in \cite{Sah2cyl}
for the region between two coaxial shells (see also
\cite{Sahrev}).

In this paper we will study the vacuum
expectation values of the energy-momentum tensor for the massive
scalar field with general curvature coupling satisfying Robin
boundary condition on the surface of a cylindrical shell in
arbitrary spatial dimensions. As for other geometries, an
important question (see, for instance, \cite{Deutsch}) is the
relation between the mode sum energy, as a renormalised sum of
the zero-point energies for each normal mode, and the volume
integral of the renormalised energy density. We will show that
for the geometry under consideration these two quantities are
different and will interpret this difference as a result of the
additional surface energy contribution (similar considerations
for the plane and spherical geometries were made in \cite{RomSah,Sahsph}).

The paper is organized as follows. Sec. 2 is devoted to the energy density
inside a cylindrical shell, from the expectation values of the
energy-momentum tensor. Calculations of this nature involve infinite sums.
The summation method applied in this case is based on a variant of the
generalized Abel-Plana formula \cite{Sahrev} together with an adequate
cutoff function. In sec. 3, the integrated energy inside the cylinder is
calculated. One finds that, in general, the total vacuum energy differs from
the integral of the volumic density. This is explained by the presence of a
surface contribution located at the boundary. The part corresponding to the
modes propagating in the exterior is considered in sec. 4. Then, it is
possible to sum both pieces and obtain a formal expression for the total
integrated Casimir energy. In sec. 5 we study the possibilities of
performing actual evaluations of this magnitude. Zeta function is proposed
as a summation method, and its working is illustrated in the example of $%
D=3+1$ and zero mass. The ensuing ending remarks are included into sec. 6

\section{\noindent Casimir energy density and stress inside a cylindrical
shell}

We will consider a scalar field $\varphi$, with curvature coupling $\xi $,
satisfying a generic Robin boundary condition
\begin{equation}
(A+Bn^{i}\nabla _{i})\varphi (x)=0  \label{robbc}
\end{equation}
on the cylindrical shell with radius $a$. Here $n^{i}$ is the
normal to the boundary surface, $\nabla _{i}$ - is the covariant
derivative operator, $A$ and $B$ are constants. The results in
the following will depend on the ratio of these coefficients
only. However, to keep the transition to the Dirichlet and
Neumann cases transparent we will use the form (\ref{robbc}). The
corresponding field equation is in form
\begin{equation}
\left( \nabla ^{i}\nabla _{i}+\xi R+m^{2}\right) \varphi (x)=0,
\label{fieldeq}
\end{equation}
where $R$ - is the scalar curvature for the background spacetime, $m$ - is
the mass for the field quanta.
By using this equation, in the case of the flat background
the corresponding metric energy-momentum tensor (EMT) may be
presented in the form
\begin{equation}
T_{ik}=\nabla _{i}\varphi \nabla _{k}\varphi +\left[ (\xi -\frac{1}{4}%
)g_{ik}\nabla ^{l}\nabla _{l}-\xi \nabla _{i}\nabla _{k} \right]
\varphi ^{2}. \label{EMT}
\end{equation}
The vacuum expectation values (v.e.v.'s) for these quantities can be derived
by evaluating the mode sum
\begin{equation}
\langle 0|T_{ik}(x)|0\rangle =\sum_{{\bf \alpha }}T_{ik}\left\{ \varphi _{%
{\bf \alpha }}(x),\varphi _{{\bf \alpha }}^{\ast }(x)\right\} .
\label{vevEMT}
\end{equation}
Here $\{\varphi _{{\bf \alpha }}(x)\}$ is a complete orthonormal set of
positive frequency solutions to the field equation with quantum numbers $%
\alpha $, satisfying the boundary condition (\ref{robbc}).

In this section we will consider the scalar vacuum inside a cylindrical
shell in $D$-dimensional spacetime. In accordance with the problem symmetry
we will take cylindrical coordinates $(r,\phi ,z_{1},\ldots ,z_{N})$, $N=D-3$%
. The corresponding eigenfunctions to (\ref {fieldeq}) have the
form
\begin{eqnarray}
\varphi _{\alpha }(x) &=&\beta _{\alpha }J_{n}(\gamma r)\exp \left(
in\phi +i{\bf kr}_{\parallel }-i\omega t\right) ,\quad \omega =\sqrt{%
\gamma ^{2}+k^{2}+m^{2}},\quad  \label{eigfunc} \\
\alpha &=&(n,\gamma ,{\bf k}),\quad n=\cdots ,-2,-1,0,1,2,\cdots ,
\end{eqnarray}
where ${\bf r}_{\parallel }=(z_{1},\ldots ,z_{N})$ and the coefficients $%
\beta _{\alpha }$ are determined from the orthonormality
condition. The eigenvalues for the quantum number $\gamma $ are
quantized by the boundary condition (\ref{robbc}) on the cylinder
surface $r=a$. From this condition it follows that the possible
values of $\gamma $ are solutions to the equation
\begin{equation}
\bar{J}_{n}(\gamma a)=AJ_{n}(\gamma a)+B\gamma J_{n}^{\prime }(\gamma a)=0.
\label{robbcin}
\end{equation}
Here and below we use the notation
\begin{equation}
\bar{f}(z)\equiv Af(z)+(B/a)zf^{\prime }(z)  \label{fbar}
\end{equation}
for a given function $f(z)$. It is well known that for real $A$ and $B$ the
all zeros for $\bar{J}_{n}(z)$ are simple and real. Let $z=\lambda _{n,l}$, $%
l=1,2,\ldots $ be the corresponding positive zeros, arranged in ascending
order: $\lambda _{n,l}<\lambda _{n,l+1}$. Now the possible values for $%
\gamma $ can be expressed as $\gamma =\lambda _{n,l}/a$.

From the orthonormality condition
\begin{equation}
\int dV\varphi _{\alpha }(x)\varphi _{\alpha ^{\prime }}^{\ast }(x)=\frac{1}{%
2\omega }\delta _{nn^{\prime }}\delta _{ll^{\prime }}\delta ({\bf k}-{\bf k}%
^{\prime }),  \label{normcond}
\end{equation}
where the integration goes over the region inside a cylinder,
one finds for the coefficients $\beta _{\alpha }$
\begin{equation}
\beta _{\alpha }^{2}=\frac{\gamma T_{n}(\gamma a)}{\omega a(2\pi )^{N+1}},
\label{betnorm}
\end{equation}
and, following \cite{Sahrev}, we have introduced the notation
\begin{equation}
T_{n}(z)=\frac{z}{(z^{2}-n^{2})J_{n}^{2}(z)+z^{2}J_{n}^{\prime
2}(z)} .\label{Tnz}
\end{equation}
Using eigenfunctions (\ref{eigfunc}) for the v.e.v. of the two-field
product (positive frequency Wightmann function) we have
\begin{eqnarray}
\langle 0|\varphi (x)\varphi (x^{\prime })|0\rangle &=&\sum_{{\bf \alpha }%
}\varphi _{{\bf \alpha }}(x)\varphi _{{\bf \alpha }}^{\ast }(x^{\prime })=%
\frac{1}{(2\pi )^{N+1}a^{2}}\sum_{{\bf \alpha }}\frac{\lambda
_{n,l}T_{n}(\lambda _{n,l})}{\sqrt{\lambda _{n,l}^{2}/a^{2}+k^{2}+m^{2}}}%
\times  \label{product} \\
&\times & J_{n}(\lambda _{n,l}r/a)J_{n}(\lambda _{n,l}r^{\prime
}/a)\exp
\left[ in(\phi -\phi ^{\prime })+i{\bf k}({\bf r}_{\parallel }-{\bf r}%
_{\parallel }^{\prime })-i\omega (t-t^{\prime })\right] ,  \nonumber
\end{eqnarray}
where
\begin{equation}
\sum_{{\bf \alpha }}=\int d^{N}{\bf k}\sum_{n=-\infty }^{+\infty
}\sum_{l=1}^{\infty }.  \label{sumnot}
\end{equation}
Substituting this expression into (\ref{vevEMT}) for the v.e.v. of
the EMT one finds
\begin{equation}
\langle 0|T_{i}^{k}|0\rangle ={\rm diag}(\varepsilon
,-p_{1},-p_{2},-p_{3},\ldots ,-p_{D-1}).  \label{emtdiag}
\end{equation}
Here $\varepsilon $ is the vacuum energy density, $p_{1}$, $p_{2}$, $%
p_{3}=p_{4}=\cdots =p_{D-1}$ are effective pressures in the radial,
azimuthal and longitudinal directions, respectively (vacuum stresses). These
quantities are determined by the relations
\begin{equation}
q(r)=\frac{1}{(2\pi )^{N+1}a^{3}}\sum_{{\bf \alpha }}\frac{\lambda
_{n,l}^{3}T_{n}(\lambda _{n,l})}{\sqrt{\lambda
_{n,l}^{2}+k^{2}a^{2}+m^{2}a^{2}}}f_{n}^{(q)}[J_{n}(\lambda
_{n,l}r/a)],\quad q=\varepsilon ,p_{i},  \label{qin}
\end{equation}
where, for a given function $f(z)$, we have introduced the notations
\begin{eqnarray}
f_{n}^{(\varepsilon )}[f(z)] &=&\left( 1+\frac{k^{2}+m^{2}}{z^{2}}%
r^{2}\right) f^{2}(z)-\left( 2\xi -\frac{1}{2}\right) \left[ f^{\prime
2}(z)+\left( \frac{n^{2}}{z^{2}}-1\right) f^{2}(z)\right]  \label{fnepsin} \\
f_{n}^{(p_{1})}[f(z)] &=&\frac{1}{2}\left[ f^{\prime 2}(z)-\left( \frac{n^{2}%
}{z^{2}}-1\right) f^{2}(z)\right] +\frac{2\xi }{z}f(z)f^{\prime }(z)
\label{fnp1in} \\
f_{n}^{(p_{2})}[f(z)] &=&\left( 2\xi -\frac{1}{2}\right) \left[ f^{\prime
2}(z)+\left( \frac{n^{2}}{z^{2}}-1\right) f^{2}(z)\right] -\frac{2\xi }{z}%
f(z)f^{\prime }(z)+\frac{n^{2}}{z^{2}}f^{2}(z)  \label{fnp2in} \\
f_{n}^{(p_{i})}[f(z)] &=&\left( 2\xi -\frac{1}{2}\right) \left[ f^{\prime
2}(z)+\left( \frac{n^{2}}{z^{2}}-1\right) f^{2}(z)\right] +\frac{k^{2}r^{2}}{%
Nz^{2}}f^{2}(z),i=3,\ldots ,D-1 . \label{fnpiin}
\end{eqnarray}
Integrating over ${\bf k}$ by using the formula
\begin{equation}
\int \frac{k^{s}d^{N}{\bf k}}{\sqrt{k^{2}+c^{2}}}=c^{N+s-1}\frac{\pi
^{(N-1)/2}}{\Gamma (N/2)}\Gamma \left( -\frac{N+s-1}{2}\right) \Gamma \left(
\frac{N+s}{2}\right)  \label{intk1}
\end{equation}
we can see that
\begin{equation}
p_{i}=-\varepsilon ,\quad i=3,\ldots ,D-1.  \label{relpieps}
\end{equation}
The v.e.v.'s (\ref{qin}) are divergent. To make them finite we introduce the
cutoff \ function $\psi _{\mu }(\gamma )$, which decreases with increasing $%
\gamma $ and satisfies the condition $\psi _{\mu }\rightarrow 1$, $\mu
\rightarrow 0$. To extract the divergent parts we will apply to the
corresponding sums over $l$ the summation formula \cite{Sahrev}
\begin{eqnarray}
\sum_{l=1}^{\infty }\frac{T_{n}(\lambda _{n,l})f(\lambda _{n,l})}{\sqrt{%
\lambda _{n,l}^{2}+c^{2}}} &=&\frac{1}{2}\int_{0}^{\infty }\frac{f(x)}{\sqrt{%
x^{2}+c^{2}}}dx-\frac{1}{2\pi }\int_{0}^{c}dx\frac{\bar{K}_{n}(x)}{\bar{I}%
_{n}(x)}\frac{e^{-n\pi i}f\left( ix\right) +e^{n\pi i}f(-ix)}{\sqrt{%
c^{2}-x^{2}}}+  \nonumber \\
&&+\frac{i}{2\pi }\int_{c}^{\infty }dx\frac{\bar{K}_{n}(x)}{\bar{I}_{n}(x)}%
\frac{e^{-n\pi i}f\left( ix\right) -e^{n\pi
i}f(-ix)}{\sqrt{x^{2}-c^{2}}}. \label{sumformula}
\end{eqnarray}
This formula is valid for functions $f(z)$ satisfying the
conditions
\begin{equation}
\left| f(z)\right| <\epsilon (x)e^{c\left| y\right| },\quad z=x+iy,\quad c<2,
\label{cond1tosum}
\end{equation}
\begin{equation}
f(z)=o(z^{2\left| n\right| -1}),\quad z\rightarrow 0,
\label{cond2tosum}
\end{equation}
where $\epsilon (x)\rightarrow 0$ for $x\rightarrow \infty $.

To evaluate the mode sum over $l$ in (\ref{qin}) in
(\ref{sumformula}) as a function $f(z)$ we choose
\begin{equation}
f(z)=z^{3}f_{n}^{(q)}[J_{n}(zr/a)]\psi _{\mu }(z/a).  \label{fztosum}
\end{equation}
We will assume a class of cutoff functions for which (\ref{fztosum})
satisfies conditions (\ref{cond1tosum}) and (\ref{cond2tosum}) uniformly
with respect to cutoff parameter $\mu $ (this is the case, for instance, for
$e^{-\mu \gamma }$, $\mu >0$). Using the relation $%
f_{n}^{(q)}[J_{n}(-izr/a)]=e^{2n\pi i}f_{n}^{(q)}[J_{n}(izr/a)]$
we see that the subintegrand of the first integral on the right hand
side of formula (\ref{sumformula}) is proportional to $\psi _{\mu
}(iz/a)-\psi _{\mu }(-iz/a)$. Consequently, after removing the
cutoff ($\psi _{\mu }\rightarrow 1$) the contribution of the first
integral will be zero. Hence, omitting this integral for the
v.e.v. we obtain
\begin{eqnarray}
q &=&\frac{1}{(2\pi )^{N+1}a^{3}}\int d^{N}{\bf k}\sum_{n=-\infty }^{+\infty
}\left\{ \frac{1}{2}\int_{0}^{\infty }dz\frac{z^{3}f_{n}^{(q)}[J_{n}(zr/a)]%
\psi _{\mu }(z/a)}{\sqrt{z^{2}+k^{2}a^{2}+m^{2}a^{2}}}+\right.
\label{qinsub1} \\
&&\left. +e^{-n\pi i}\frac{1}{\pi }\int_{a\sqrt{k^{2}+m^{2}}}^{\infty }dz%
\frac{\bar{K}_{n}(z)}{\bar{I}_{n}(z)}\frac{z^{3}f_{n}^{(q)}[J_{n}(izr/a)]%
\chi _{\mu }(z/a)}{\sqrt{z^{2}-k^{2}a^{2}-m^{2}a^{2}}}\right\} ,  \nonumber
\end{eqnarray}
where $\chi _{\mu }(y)=[\psi _{\mu }(iy)+\psi _{\mu }(-iy)]/2$. The second
integral on the right of this formula vanishes in the limit $a\rightarrow
\infty $, whereas the first one does not depend on $a$. It follows from here
that the latter corresponds to the spacetime without boundaries. This can be
also seen directly by explicit summation over $n$ using the formula
$\ds\sum_{n=-\infty }^{+\infty }J_{n\pm p}^{2}(z)=1$. For instance, in the case
of the energy density it follows from here that
\begin{equation}
\sum_{n=-\infty }^{+\infty }f_{n}^{(\varepsilon )}[J_{n}(zr)]=1+\frac{%
k^{2}+m^{2}}{z^{2}}.  \label{fnepssum}
\end{equation}
Using this relation for the term corresponding to the first integral on the
right of (\ref{qinsub1}) and removing the cutoff one has
\begin{eqnarray}
\varepsilon ^{(0)} &=&\frac{1}{2(2\pi )^{N+1}}\int d^{N}{\bf k}%
\sum_{n=-\infty }^{+\infty }\int_{0}^{\infty }dz\frac{%
z^{3}f_{n}^{(\varepsilon )}[J_{n}(zr)]\psi _{\mu
}(z/a)}{\sqrt{z^{2}+k^{2}+m^{2}}}=
\label{epsMink} \\
&=&\frac{1}{2(2\pi )^{N+1}}\int d^{N}{\bf k}\int_{0}^{\infty }dz\,z\sqrt{%
z^{2}+k^{2}+m^{2}}=\frac{1}{2}\int \frac{d^{D-1}{\bf k}}{(2\pi )^{D-1}}\sqrt{%
k^{2}+m^{2}}.  \nonumber
\end{eqnarray}
Hence summation formula (\ref{sumformula}) allows us to extract the
contribution of the unbounded space without specifying the cutoff function.
In the remaining part, the integration over ${\bf k}$ can be explicitly done
using the formula
\begin{equation}
\int d^{N}{\bf k}\int_{\sqrt{k^{2}+m^{2}}}^{\infty }\,\frac{k^{s}g(z)dz}{%
\sqrt{z^{2}-k^{2}-m^{2}}}=\frac{\pi ^{N/2}}{\Gamma (N/2)}B\left( \frac{N+s}{2%
},\frac{1}{2}\right) \int_{m}^{\infty }dz\,\left( z^{2}-m^{2}\right)
^{(N+s-1)/2}g(z),  \label{intk2}
\end{equation}
where $B(x,y)$ is the Euler beta function (this formula can be
proved by integrating over the angular part of $d^{N}{\bf k}$ and
introducing polar coordinates in the plane
$(x=\sqrt{z^{2}-k^{2}-m^{2}},k)$). Removing the cutoff for this
part one obtains
\begin{equation}
q_{SUB}=q-q^{(0)}=\frac{2^{2-D}\pi ^{-D/2}}{a^{D}\Gamma (D/2-1)}%
\sum_{n=-\infty }^{+\infty }\int_{ma}^{\infty }dz\,z^{3}\left(
z^{2}-a^{2}m^{2}\right) ^{D/2-2}\frac{\bar{K}_{n}(z)}{\bar{I}_{n}(z)}%
F_{n}^{(q)}[I_{n}(zr/a)],  \label{qinsub}
\end{equation}
where $I_{n}(z)$ and $K_{n}(z)$ are the modified Bessel functions, and
\begin{eqnarray}
F_{n}^{(\varepsilon )}[f(z)] &=&\frac{1}{D-2}\left( 1-\frac{m^{2}}{z^{2}}%
r^{2}\right) f^{2}(z)+\left( 2\xi -\frac{1}{2}\right) \left[ f^{\prime
2}(z)+\left( \frac{n^{2}}{z^{2}}+1\right) f^{2}(z)\right]  \label{Fnepsin} \\
F_{n}^{(p_{1})}[f(z)] &=&\frac{1}{2}\left[ \left( \frac{n^{2}}{z^{2}}%
+1\right) f^{2}(z)-f^{\prime 2}(z)\right] -\frac{2\xi }{z}f(z)f^{\prime }(z)
\label{Fnp1in} \\
F_{n}^{(p_{2})}[f(z)] &=&-\left( 2\xi -\frac{1}{2}\right) \left[ f^{\prime
2}(z)+\left( \frac{n^{2}}{z^{2}}+1\right) f^{2}(z)\right] +\frac{2\xi }{z}%
f(z)f^{\prime }(z)-\frac{n^{2}}{z^{2}}f^{2}(z)  \label{Fnp2in} \\
F_{n}^{(p_{i})}[f(z)] &=&-F_{n}^{(\varepsilon )}[f(z)],\quad i=3,\ldots ,D-1.
\label{Fnpiin}
\end{eqnarray}
In (\ref{qinsub}) $q^{(0)}$ are the components of the v.e.v.
$\langle 0_{M}|T_{i}^{k}|0_{M}\rangle $ over the vacuum
$|0_{M}\rangle $\ corresponding to the unbounded $D$-dimensional
Minkowskian spacetime. It can be seen that v.e.v.'s (\ref{qinsub})
satisfy the continuity equation $\nabla _{i}T_{k}^{i}=0$, which,
for the cylindrical geometry, takes the form
\begin{equation}
\frac{dp_{1}}{dr}+\frac{1}{r}\left( p_{1}-p_{2}\right) =0.  \label{conteqcyl}
\end{equation}
For $D=3$ formulae (\ref{qinsub}) coincide with the corresponding
results in \cite{Sahsph} for the spherical case.

Formulae (\ref{qinsub}) can be obtained by another equivalent
way, applying a certain second-order differential operator on the
regularized Wightmann function and taking the coincidence limit.
The calculation of the Wightmann function is of interest for its
own sake, as this function determines the response of a particle
detector moving through the vacuum under consideration
\cite{Birrel}. For this reason here we derive formula for the
regularized Wightmann function. To do this we apply summation
formula (\ref{sumformula}) to the sum over $l$ in
(\ref{product}). It can be shown that the first integral on the
right of (\ref{sumformula}) will give the corresponding Wightmann
function for $D$-dimensional Minkowskian spacetime, and the
second integral vanishes. Integrating over ${\bf k}$ with the
help of standard integrals, in the term corresponding to the third
integral one obtains
\begin{eqnarray}
\langle 0\vert \varphi (x)\varphi (x')\vert 0\rangle &=& \langle
0_{M}\vert \varphi (x)\varphi (x')\vert 0_{M}\rangle -
\sum_{n=-\infty}^{+\infty }\frac{e^{in(\phi -\phi ')}}{(2\pi
)^{D/2}}\int_{m}^{\infty }dz\, z(z^2-m^2)^{D/4-1}\times \\
&\times & I_n(zr)I_n(zr')\frac{\bar K_{n}(z)}{\bar
I_{n}(z)}\frac{J_{D/2-2}\left( \sqrt{z^2-m^2}\left[ ({\bf
r}_{\parallel }-{\bf r}'_{\parallel })^2-(t-t')^2 \right] ^{1/2}
\right)}{\left[ ({\bf r}_{\parallel }-{\bf r}'_{\parallel
})^2-(t-t')^2 \right] ^{D/4-1}}. \nonumber \label{Wightregcyl}
\end{eqnarray}
It can be easily seen that for the case of a circle ($D=3$) this
formula coincides with the corresponding expression for the
spherical case derived in \cite{Sahsph}.

\begin{figure}[tbph]
\begin{center}
\begin{tabular}{ccc}
\epsfig{figure=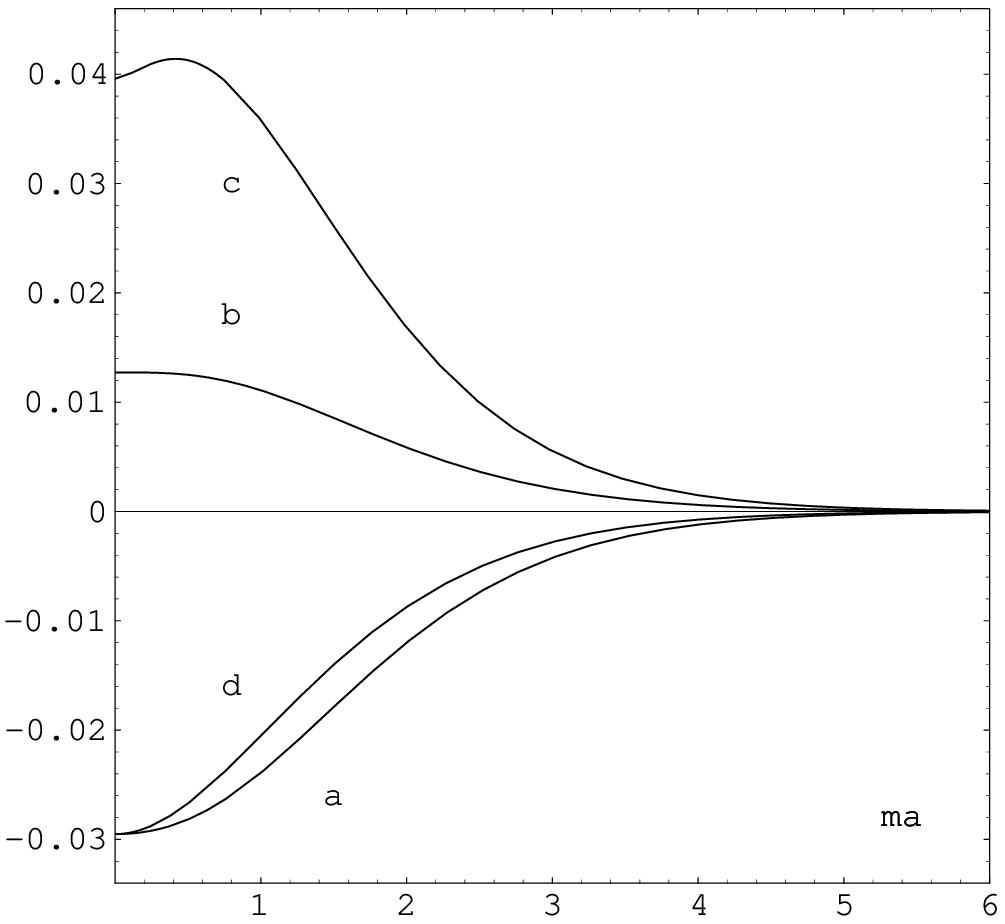,width=7cm,height=7cm} & \hspace*{0.5cm} & %
\epsfig{figure=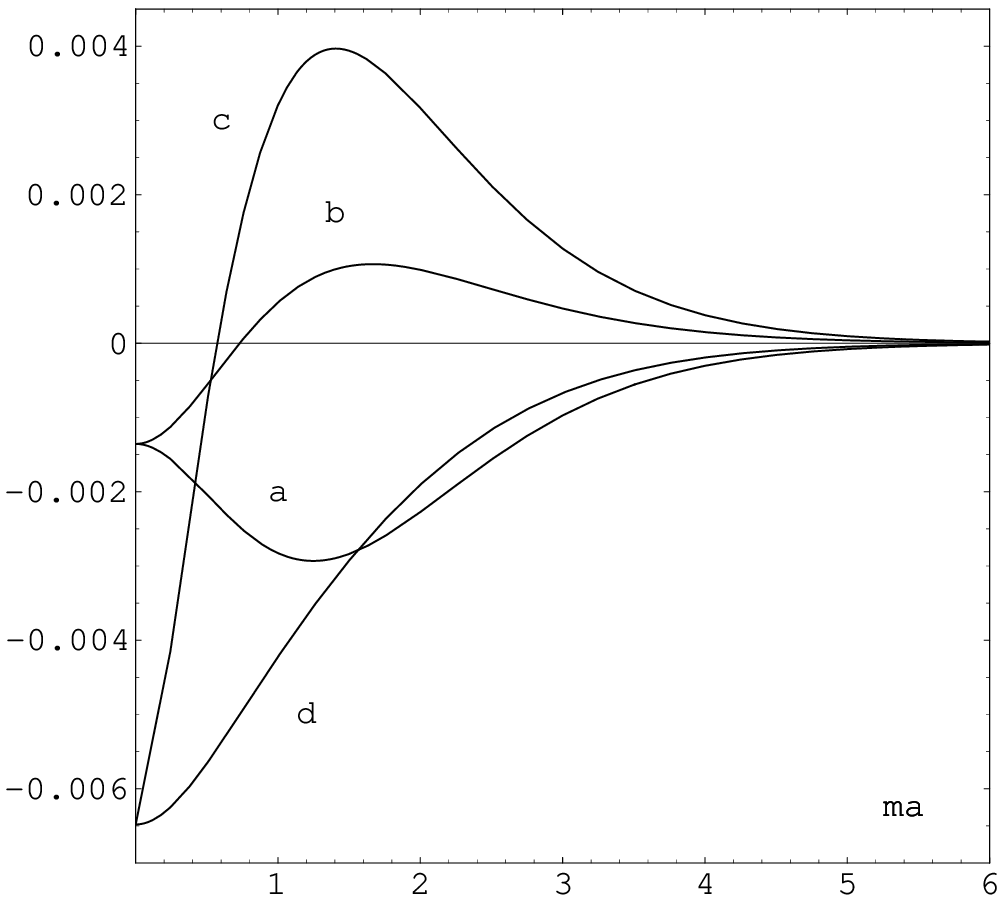,width=7cm,height=7cm}
\end{tabular}
\end{center}
\caption{ The Casimir energy density, $a^{D}\protect\varepsilon
$, (a, c) and vacuum radial pressure, $a^{D}p_{1}$, (b, d) on the
cylinder axis for minimally (left) and conformally (right)
coupled Dirichlet (a, b) and Neumann (c, d) scalars in $D=4$ as
functions of $ma$.} \label{figcylax}
\end{figure}

 Let's consider vacuum densities
(\ref{qinsub}) on the cylinder axis, $r=0$. Now the only
contributions come from the summands with $n=0,\pm 1$ and one has
\begin{equation}
q(r=0)=\frac{2^{2-D}\pi ^{-D/2}}{a^{D}\Gamma (D/2-1)}\int_{ma}^{\infty
}dz\,z^{3}\left( z^{2}-a^{2}m^{2}\right) ^{D/2-2}\left\{ u_{q0}\frac{\bar{K}%
_{0}(z)}{\bar{I}_{0}(z)}+u_{q1}\frac{\bar{K}_{1}(z)}{\bar{I}_{1}(z)}\right\}
,  \label{qcentre}
\end{equation}
where
\begin{eqnarray}
u_{\varepsilon 0} &=&\frac{1-m^{2}a^{2}/z^{2}}{D-2}+2\xi -\frac{1}{2}%
,\;u_{\varepsilon 1}=2\xi -\frac{1}{2}  \label{uaxes} \\
u_{p_{1}0} &=&u_{p_{2}0}=-\xi +\frac{1}{2},\qquad u_{p_{1}1}=u_{p_{2}1}=-\xi
.  \nonumber
\end{eqnarray}
The results of the corresponding numerical evaluation for the Dirichlet and
Neumann minimally and conformally coupled scalars in $D=4$ are presented in
Fig.\ref{figcylax}.

In the large mass limit, $ma\gg 1$, using the asymptotic formulae for the
modified Bessel function for large values of argument in the leading order
from (\ref{qcentre}) one obtains
\begin{equation}
\varepsilon (r=0)\sim -2p_{1}(r=0)\sim \frac{4\xi -1}{2^{D-1}\pi
^{D/2-1}a^{D}}(am)^{D/2+1}e^{-2am}\left( 2\delta _{B0}-1\right) ,\;ma\gg 1.
\label{axesmass}
\end{equation}
The v.e.v.'s (\ref{qinsub}) diverge on the cylinder surface (for a
given $n$ the $z$-integrals diverge as $(a-r)^{1-D}$, $r\to a$).
The corresponding asymptotic behaviour can be found using the
uniform asymptotic expansions for the modified Bessel functions,
and the leading terms have the form
\begin{eqnarray}
\varepsilon &\sim &-p_{2}\sim \frac{(D-1)(\xi -\xi _{c})\Gamma (D/2)}{%
2^{D-1}\pi ^{D/2}(a-r)^{D}}\left( 2\delta _{B0}-1\right)  \label{nearsurfin}
\\
p_{1} &\sim &-\frac{(\xi -\xi _{c})\Gamma (D/2)}{2^{D-1}\pi
^{D/2}a(a-r)^{D-1}}\left( 2\delta _{B0}-1\right) .  \nonumber
\end{eqnarray}
Note that these terms are independent of the mass and Robin
coefficients. Taking the limit $a,r\rightarrow \infty $, $a-r={\rm
const}$ one obtains the leading terms for the asymptotic
behaviour near the single plate considered in ref.\cite{RomSah}. The
surface divergences in the renormalized expectation values for
the vacuum EMT are well known in quantum field theory. In the
case of $D=3$ massless fields the corresponding asymptotic series
near arbitrary smooth boundary are presented in refs.\cite{Deutsch,KCD}.

\section{Total Casimir energy inside a cylinder}

The unregularized integrated Casimir energy inside a cylindrical
shell may be obtained by integrating (\ref{qin}) for
$q=\varepsilon $. Using standard formulae for the integrals
involving the Bessel functions this energy can be presented in the
form
\begin{equation}
E_{{\rm in}}^{{\rm (vol)}}=E_{{\rm in}}-\left( 2\xi -\frac{1}{2}\right)
\left( \frac{L}{2\pi }\right) ^{N}\frac{1}{a}\sum_{\alpha }\frac{\lambda
_{n,l}^{2}T_{n}(\lambda _{n,l})}{\sqrt{\lambda
_{n,l}^{2}+k^{2}a^{2}+m^{2}a^{2}}}J_{n}(\lambda _{n,l})J_{n}^{\prime
}(\lambda _{n,l}),  \label{Evolin}
\end{equation}
where
\begin{equation}
E_{{\rm in}}=\frac{1}{2}\left( \frac{L}{2\pi }\right) ^{N}\sum_{\alpha }%
\sqrt{\lambda _{n,l}^{2}/a^{2}+k^{2}+m^{2}}  \label{Ein}
\end{equation}
is the total Casimir energy inside a cylinder, evaluated as a sum
of the zero-point energies for each normal mode of frequency. As
we see, for the general Robin case, (\ref{Evolin}) differs from
the total vacuum energy. The reason for this difference is the
existence of an additional surface energy contribution to the
volume energy, located on the boundary $r=a-0$:
\begin{equation}
E_{{\rm in}}^{{\rm (surf)}}=\left( 2\xi -\frac{1}{2}\right) \left( \frac{L}{%
2\pi }\right) ^{N}\frac{1}{a}\sum_{\alpha }\frac{\lambda
_{n,l}^{2}T_{n}(\lambda _{n,l})}{\sqrt{\lambda
_{n,l}^{2}+k^{2}a^{2}+m^{2}a^{2}}}J_{n}(\lambda _{n,l})J_{n}^{\prime
}(\lambda _{n,l}).  \label{Esurfin}
\end{equation}
To see this, note that there is a surface energy density contribution to
the vacuum energy density \cite{KCD}:
\begin{equation}
T_{00}^{{\rm (surf)}}=-\left( 2\xi -\frac{1}{2}\right) \delta (r-a+0)\varphi
n^{i}\partial _{i}\varphi .  \label{Tsurfin}
\end{equation}
The corresponding v.e.v. can be evaluated using eigenfunctions
 (\ref{eigfunc}):
\begin{equation}
\langle 0|T_{00}^{{\rm (surf)}}|0\rangle =\delta (r-a+0)\frac{2\xi -1/2}{%
a^{2}(2\pi )^{N}}\sum_{\alpha }\frac{\lambda _{n,l}^{2}T_{n}(\lambda _{n,l})%
}{\sqrt{\lambda _{n,l}^{2}+k^{2}a^{2}+m^{2}a^{2}}}J_{n}(\lambda
_{n,l}r/a)J_{n}^{\prime }(\lambda _{n,l}r/a).  \label{vevTsurfin}
\end{equation}
The integration of this energy density leads to the expression (\ref{Esurfin}%
) for the total interior surface energy. In (\ref{vevTsurfin}), applying to
the sum over $l$ formula (\ref{sumformula}), in analogy to (\ref{qinsub}),
the subtracted surface energy density can be presented in the form
\begin{eqnarray}
\left\langle T_{00}^{{\rm (surf)}}\right\rangle _{SUB} &=&-\frac{\left( 2\xi
-1/2\right) \delta (r-a+0)}{2^{D-2}\pi ^{D/2}\Gamma (D/2-1)a^{D-1}}%
\sum_{n=-\infty }^{+\infty }\int_{ma}^{\infty }dz\,z^{2}\left(
z^{2}-m^{2}a^{2}\right) ^{D/2-2}\times  \nonumber \\
&&\times \frac{\bar{K}_{n}(z)}{\bar{I}_{n}(z)}I_{n}\left( z\frac{r}{a}%
\right) I_{n}^{\prime }\left( z\frac{r}{a}\right) .  \label{Tsurfsub}
\end{eqnarray}
This leads to the following subtracted surface energy
\begin{equation}
\left\langle E_{{\rm in}}^{{\rm (surf)}}\right\rangle _{SUB}=\frac{-\left(
4\xi -1\right) L^{D-3}a^{2-D}}{2^{D-2}\pi ^{D/2-1}\Gamma (D/2-1)}%
\sum_{n=-\infty }^{+\infty }\int_{ma}^{\infty }dz\,z^{2}\left(
z^{2}-m^{2}a^{2}\right) ^{D/2-2}\frac{\bar{K}_{n}(z)}{\bar{I}_{n}(z)}%
I_{n}\left( z\right) I_{n}^{\prime }\left( z\right) .  \label{Esurfinsub}
\end{equation}
We can find the subtracted total volume energy inside a cylinder integrating
the energy density (\ref{qinsub}):
\begin{equation}
\left\langle E_{{\rm in}}^{{\rm (vol)}}\right\rangle _{SUB}=\frac{%
L^{D-3}a^{2-D}}{2^{D-2}\pi ^{D/2-1}\Gamma (D/2-1)}\sum_{n=-\infty
}^{+\infty }\int_{ma}^{\infty }dz\,z\left( z^{2}-m^{2}a^{2}\right)
^{D/2-2}\frac{\bar{K}_{n}(z)}{\bar{I}_{n}(z)}F_{{\rm
v}}[I_{n}(z)], \label{Evolinsub}
\end{equation}
where, for a given function $f(z)$, we have introduced the notation
\begin{equation}
F_{{\rm v}}[f(z)]=(4\xi -1)zf(z)f^{\prime }(z)-\frac{z^{2}-m^{2}a^{2}}{D-2}%
\left[ f^{\prime 2}(z)-\left( 1+\frac{n^{2}}{z^{2}}\right) f^{2}(z)\right] .
\label{Fv}
\end{equation}
The total interior vacuum energy can be obtained as a sum of the surface and
volume parts:
\begin{eqnarray}
\left\langle E_{{\rm in}}\right\rangle _{SUB}
&=&-\frac{L^{D-3}a^{2-D}}{2^{D-1}\pi ^{D/2-1}\Gamma
(D/2)}\sum_{n=-\infty }^{+\infty }\int_{ma}^{\infty
}dz\,z\left( z^{2}-m^{2}a^{2}\right) ^{D/2-1}\times  \label{Einsub} \\
&\times & \frac{\bar{K}_{n}(z)}{\bar{I}_{n}(z)}\left[
I_{n}^{\prime 2}(z)-\left( 1+\frac{n^{2}}{z^{2}}\right)
I_{n}^{2}(z)\right] . \nonumber
\end{eqnarray}
As we see, in the total vacuum energy the dependencies on the
curvature coupling $\xi $ are cancelled out and we obtain $\xi
$-independent vacuum energy. We could have expected this result, as
eigenfrequencies are independent of this parameter. The vacuum
force acting from inside per unit surface of the cylinder can be
found using the expression (\ref{qinsub}) for the vacuum radial
pressure:
\begin{equation}
F_{{\rm in}}=p_{1}\mid _{r=a-0}=\frac{2^{2-D}\pi ^{-D/2}}{a^{D}\Gamma (D/2-1)%
}\sum_{n=-\infty }^{+\infty }\int_{ma}^{\infty }dz\,z^{3}\left(
z^{2}-m^{2}a^{2}\right) ^{D/2-2}\frac{\bar{K}_{n}(z)}{\bar{I}_{n}(z)}%
F_{n}^{(p_{1})}[I_{n}(z)],  \label{Fin}
\end{equation}
with notation (\ref{Fnp1in}). For the massless scalar field and
in the case of conformal coupling, $\xi =\xi _{c}$, the vacuum
force and volume energy are related by
\begin{equation}
\left\langle E_{{\rm in}}^{{\rm (vol)}}\right\rangle _{SUB}=
\frac{2\pi L^{D-3}%
}{D-2}a^{2}F_{{\rm in}}.  \label{EvolFin}
\end{equation}
This relation can be also obtained from the continuity equation (\ref
{conteqcyl}) if we take into account the tracelessness condition for the
corresponding energy-momentum tensor.

\section{Casimir densities outside a cylindrical shell}

Firstly, let's consider v.e.v.'s for the energy-momentum tensor in the
region between two coaxial cylindrical layers with radii $a$ and $b$, $a<b$.
The corresponding boundary conditions are in form
\begin{equation}
\left( A+B\frac{\partial }{\partial r}\right) \varphi (x)=0,\quad
r=a,b . \label{bc2cyl}
\end{equation}
The corresponding eigenfunctions can be obtained from
(\ref{eigfunc}) with the replacement
\begin{equation}
J_{n}(\gamma r)\rightarrow g_{n}(\gamma a,\gamma r)=J_{n}(\gamma r)\bar{%
Y}_{n}(\gamma a)-Y_{n}(\gamma r)\bar{J}_{n}(\gamma a),  \label{gn}
\end{equation}
where $Y_{n}(z)$ is the Neumann function. The functions chosen in
this way satisfy the boundary condition on the inner cylinder.
From the boundary
condition on the shell $r=b$ one obtains that the possible values of $%
\gamma $ are solutions to the equation
\begin{equation}
C_{n}^{ab}(\eta ,\gamma a)\equiv \bar{J}_{n}(\gamma a)\bar{Y}%
_{n}(\gamma b)-\bar{Y}_{n}(\gamma a)\bar{J}_{n}(\gamma b)=0.
\label{Cnab}
\end{equation}
The corresponding roots will be denoted by $\gamma a=\sigma _{n,l}$, $%
l=1,2,\ldots $.

The normalization coefficient is determined from condition (\ref{normcond}%
), where now the integration goes over the region between the
cylindrical shells, and is equal to
\begin{equation}
\beta _{\alpha }^{2}=\frac{\pi ^{2}\gamma T_{n}^{ab}(\gamma a)}{4\omega
a(2\pi )^{N+1}}.  \label{betnorm2cyl}
\end{equation}
Here we use the notation
\begin{equation}
T_{n}^{ab}(z)=z\left\{ \frac{\bar{J}_{n}^{2}(z)}{\bar{J}%
_{n}^{2}(\eta z)}\left[ A^{2}+B^{2}(\eta ^{2}z^{2}-n^{2})\right]
-A^{2}-B^{2}(z^{2}-n^{2})\right\} ^{-1},\quad \eta =\frac{b}{a}.
\label{Tn2cyl}
\end{equation}
Using the corresponding eigenfunctions it can be seen that the
v.e.v.'s for the energy-momentum tensor have the form
(\ref{emtdiag}), where the components are given by formulae
\begin{equation}
q(r)=\frac{\pi ^{2}}{4(2\pi )^{N+1}a^{3}}\sum_{{\bf \alpha
}}\frac{\sigma _{n,l}^{3}T_{n}^{ab}(\sigma _{n,l})}{\sqrt{\sigma
_{n,l}^{2}+k^{2}a^{2}+m^{2}a^{2}}}f_{n}^{(q)}[g_{n}(\sigma
_{n,l},\sigma _{n,l}r/a)],\quad q=\varepsilon ,p_{i},
\label{q2cyl}
\end{equation}
with the functions $f_{n}^{(q)}[f(z)]$ defined as
(\ref{fnepsin})-(\ref {fnpiin}).

To evaluate the sum over $l$ we will apply the summation formula \cite
{Sahrev}
\begin{eqnarray}
\frac{\pi ^{2}}{2}\sum_{l=1}^{\infty }h(\sigma
_{n,k})T_{n}^{ab}(\sigma
_{n,k}) &=&\int_{0}^{\infty }\frac{h(x)dx}{\bar{J}_{n}^{2}(x)+\bar{%
Y}_{n}^{2}(x)}-  \label{sumfortwocyl} \\
& - & \frac{\pi }{4}\int_{0}^{\infty }\frac{\bar{K}_{n}(\eta x)}{\bar{%
K}_{n}(x)}\frac{\left[ h(xe^{\pi i/2})+h(xe^{-\pi i/2})\right] dx}{\bar{%
K}_{n}(x)\bar{I}_{n}(\eta x)-\bar{K}_{n}(\eta x)\bar{I}_{n}(x)}{,}
\nonumber
\end{eqnarray}
where we have assumed that all zeros for the function $C_{n}^{ab}(\eta
,z)$ are real. In case of existence of purely imaginary zeros we
have to include additional residue terms on the left of formula
(\ref{sumfortwocyl}). Now, by an evaluation similar to
(\ref{qinsub1}), one can see that
\begin{eqnarray}
q &=&\frac{1}{(2\pi )^{D-2}a^{3}}\int d^{N}{\bf k}\sum_{n=-\infty
}^{+\infty
}\left\{ \frac{1}{2}\int_{0}^{\infty }\frac{dz}{\bar{J}_{n}^{2}(z)+%
\bar{Y}_{n}^{2}(z)}\frac{z^{3}f_{n}^{(q)}[g_{n}(z,zr/a)]}{\sqrt{%
z^{2}+k^{2}a^{2}+m^{2}a^{2}}}+\right.  \label{q2cyl1} \\
&&\left. +\frac{\pi }{4}\int_{a\sqrt{k^{2}+m^{2}}}^{\infty }\frac{%
z^{3}f_{n}^{(q)}[g_{n}(iz,izr/a)]}{\sqrt{z^{2}-k^{2}a^{2}-m^{2}a^{2}}}\frac{%
\bar{K}_{n}(\eta x)/\bar{K}_{n}(x)dz}{\bar{K}_{n}(x)\bar{%
I}_{n}(\eta x)-\bar{K}_{n}(\eta x)\bar{I}_{n}(x)}\right\} .  \nonumber
\end{eqnarray}
To obtain v.e.v.'s for the exterior of a single cylindrical shell let's
consider the limit $b\rightarrow \infty $. It can be easily seen that in
this limit the second integral on the right of formula (\ref{q2cyl1}) tends
to zero, whereas the first one is independent from $b$. It follows from here
that the expressions
\begin{equation}
q=\frac{1}{2(2\pi )^{D-2}a^{3}}\int d^{N}{\bf k}\sum_{n=-\infty
}^{+\infty
}\int_{0}^{\infty }\frac{dz}{\bar{J}_{n}^{2}(z)+\bar{Y}_{n}^{2}(z)}%
\frac{z^{3}f_{n}^{(q)}[g_{n}(z,zr/a)]}{\sqrt{z^{2}+k^{2}a^{2}+m^{2}a^{2}}}
\label{qoutunreg}
\end{equation}
are v.e.v.'s for the EMT components for the exterior region of a single
shell. To regularize this quantities we have to subtract the parts
corresponding to the unbounded space. As we saw, the latter can be presented
in the form (\ref{epsMink}). Using the identity
\begin{equation}
\frac{f_{n}^{(q)}[g_{n}(z,zr/a)]}{\bar{J}_{n}^{2}(z)+\bar{Y}%
_{n}^{2}(z)}-f_{n}^{(q)}[J_{n}(zr/a)]=-\frac{1}{2}\sum_{\sigma =1}^{2}\frac{%
\bar{J}_{n}(z)}{\bar{H}_{n}^{(\sigma )}(z)}f_{n}^{(q)}[H_{n}^{(%
\sigma )}(zr/a)]  \label{relcylout}
\end{equation}
with $H_{n}^{(\sigma )}(z)$, $\sigma =1,2$ being the Hankel
functions, one obtains
\begin{equation}
q=-\frac{1}{4(2\pi )^{D-2}a^{3}}\int d^{N}{\bf k}\sum_{n=-\infty
}^{+\infty
}\sum_{\sigma =1}^{2}\int_{0}^{\infty }dz\frac{\bar{J}_{n}(z)}{%
\bar{H}_{n}^{(\sigma )}(z)}\frac{z^{3}f_{n}^{(q)}[H_{n}^{(\sigma
)}(zr/a)]}{\sqrt{z^{2}+k^{2}a^{2}+m^{2}a^{2}}}.  \label{qoutsub1}
\end{equation}
Assuming that the function $\bar H_{n}^{(1)}(z)$, ($\bar
H_{n}^{(2)}(z)$) has no zeros for $0<{\rm arg}\,z\leq \pi /2$
($-\pi /2\leq {\rm arg}\,z<0$) we can rotate the integration
contour for $z$ by angle $\pi /2$ for $\sigma
=1$ and by angle $-\pi /2$ for $\sigma =2$. The integrals
over $(0,ia\sqrt{%
k^{2}+m^{2}})$ and $(0,-ia\sqrt{k^{2}+m^{2}})$ cancel out.
Introducing the Bessel modified functions and integrating over
${\bf k}$ with the help of formula (\ref{intk2}) for the
subtracted v.e.v. one obtains
\begin{equation}
q_{SUB}=\frac{2^{2-D}\pi ^{-D/2}}{a^{D}\Gamma (D/2-1)}\sum_{n=-\infty
}^{+\infty }\int_{ma}^{\infty }dz\,z^{3}\left( z^{2}-a^{2}m^{2}\right)
^{D/2-2}\frac{\bar{I}_{n}(z)}{\bar{K}_{n}(z)}%
F_{n}^{(q)}[K_{n}(zr/a)],  \label{qsubout}
\end{equation}
where we use notations (\ref{Fnepsin})-(\ref{Fnpiin}). As we see,
these quantities can be obtained from the ones for interior
region by the replacements $I\rightarrow K$, $K\rightarrow I$. As
for the interior region the v.e.v. (\ref{qsubout}) diverge at
cylinder surface. The leading terms of these divergences are
determined by the same formulae (\ref{nearsurfin}) with replacement
$a-r\rightarrow r-a$.

Let's consider the asymptotic behaviour of v.e.v (\ref{qsubout})
at large distances, $r\rightarrow \infty $, for the massless
case. Introducing a new integration variable $y=zr/a$, and
expanding the subintegral in terms of $r/a$, we can see that the
leading term of the asymptotic expansion comes from the summand
with $n=0$:
\begin{equation}
q_{SUB}=\frac{2^{2-D}\pi ^{-D/2}}{\Gamma (D/2-1)r^{D}\ln r/a}%
\int_{0}^{\infty }dy\,y^{D-1}F_{0}^{(q)}[K_{0}(y)],\qquad r\rightarrow
\infty .  \label{qfar}
\end{equation}
The integral in this expression can be evaluated using the formula for the
integrals containing the square of the McDonald function (see, for instance,
\cite{PrudnikovSF}). This leads to the following result
\begin{equation}
\varepsilon \sim (D-2)p_{1}\sim -\frac{D-2}{D-1}p_{2}\sim \frac{\Gamma
^{3}(D/2)}{\pi ^{D/2}\Gamma (D-1)}\frac{\xi -\xi _{c}}{r^{D}\ln r/a}%
,\;r\rightarrow \infty .  \label{qfar1}
\end{equation}
For a conformally coupled scalar this leading term is zero and
$\varepsilon \sim 1/r^{D+2}$.

Integrating the vacuum energy density over the region outside of
a cylindrical shell we obtain the corresponding volume energy:
\begin{equation}
\left\langle E_{{\rm ext}}^{{\rm (vol)}}\right\rangle _{SUB}=\frac{%
-L^{D-3}a^{2-D}}{2^{D-2}\pi ^{D/2-1}\Gamma
(D/2-1)}\sum_{n=-\infty }^{+\infty
}\int_{ma}^{\infty }dz\,z \left( z^{2}-m^{2}a^{2}\right) ^{D/2-2}\frac{%
\bar{I}_{n}(z)}{\bar{K}_{n}(z)}F_{{\rm v}}[K_{n}(z)],
\label{Evolext}
\end{equation}
where the functional $F_{{\rm v}}[f(z)]$ is defined as (\ref{Fv}). The
outside surface energy can be derived using the surface energy density given
by the formula (\ref{Tsurfin}) with the replacement $a-0\rightarrow a+0$ in
the argument of the $\delta $-function. This gives
\begin{equation}
\left\langle E_{{\rm ext}}^{{\rm (surf)}}\right\rangle _{SUB}=\frac{\left(
4\xi -1\right) L^{N}a^{2-D}}{2^{D-2}\pi ^{D/2-1}\Gamma (D/2-1)}%
\sum_{n=-\infty }^{+\infty }\int_{ma}^{\infty }dz\,z^{2}\left(
z^{2}-m^{2}a^{2}\right) ^{D/2-2}\frac{\bar{I}_{n}(z)}{\bar{K}%
_{n}(z)}K_{n}\left( z\right) K_{n}^{\prime }\left( z\right) .
\label{Esurfext}
\end{equation}
Now taking the sum of (\ref{Evolext}) and (\ref{Esurfext}) for
the total exterior vacuum energy one obtains
\begin{eqnarray}
\left\langle E_{{\rm ext}}\right\rangle _{SUB} &=&
\frac{L^{D-3}a^{2-D}}{2^{D-1}\pi ^{D/2-1}\Gamma
(D/2)}\sum_{n=-\infty }^{+\infty }\int_{ma}^{\infty
}dz\,z\left( z^{2}-m^{2}a^{2}\right) ^{D/2-1}\times  \label{Eextsub} \\
&\times & \frac{\bar{I}_{n}(z)}{\bar{K}_{n}(z)}\left[
K_{n}^{\prime 2}(z)-\left( 1+\frac{n^{2}}{z^{2}}\right)
K_{n}^{2}(z)\right] . \nonumber
\end{eqnarray}
As we could have expected, the dependencies on the curvature coupling are
cancelled. The expression for the radial projection of the vacuum
force acting per unit
surface of the cylinder from the outside directly follows from (\ref{qsubout}%
) with $q=p_{1}$:
\begin{equation}
F_{{\rm ext}}=-p_{1}\mid _{r=a+0}=\frac{-2^{2-D}\pi ^{-D/2}}{a^{D}\Gamma
(D/2-1)}\sum_{n=-\infty }^{+\infty }\int_{ma}^{\infty }dz\,z^{3}\left(
z^{2}-m^{2}a^{2}\right) ^{D/2-2}\frac{\bar{I}_{n}(z)}{\bar{K}%
_{n}(z)}F_{n}^{(p_{1})}[K_{n}(z)],  \label{Fext}
\end{equation}
where the function $F_{n}^{p_{1}}$ is defined as (\ref{Fnp1in}).
Now we turn to the case of a cylindrical shell with zero
thickness, assuming that the coefficients in the Robin boundary
conditions are the same for the interior and exterior regions.
The total vacuum energies end force per unit surface can be
obtained summing the corresponding quantities for these regions:
\begin{equation}
\left\langle E\right\rangle _{SUB}=\left\langle E_{{\rm in}}\right\rangle
_{SUB}+\left\langle E_{{\rm ext}}\right\rangle _{SUB},\quad F=F_{{\rm in}%
}+F_{{\rm ext}}.  \label{EFinext}
\end{equation}
Using the expressions for the interior and exterior quantities we have
\begin{eqnarray}
\left\langle E^{{\rm (surf)}}\right\rangle _{SUB}&=&\frac{\left(
4\xi -1\right) L^{D-3}a^{2-D}}{2^{D-2}\pi ^{D/2-1}\Gamma
(D/2-1)}\sum_{n=-\infty
}^{+\infty }\int_{ma}^{\infty }dz\,z\left( z^{2}-m^{2}a^{2}\right) ^{D/2-2}%
\times \nonumber \\
& & \times \left[ 1-\beta \frac{\left(
\tilde{I}_{n}(z)\tilde{K}_{n}(z)\right)
^{\prime }}{z\tilde{I}_{n}^{\prime }(z)\tilde{K}_{n}^{\prime }(z)}%
\right] ,  \label{Esurfgum}
\end{eqnarray}
for the surface part of the vacuum energy,
\begin{eqnarray}
\left\langle E\right\rangle _{SUB}&=&
\frac{-L^{D-3}a^{2-D}}{2^{D-1}\pi
^{D/2-1}\Gamma (D/2)}\sum_{n=-\infty }^{+\infty }\int_{ma}^{\infty }dz\,%
\frac{\left( z^{2}-m^{2}a^{2}\right) ^{D/2-1}}{z} \times
\nonumber \\
& & \times \left[ 2\beta
+(z^{2}+n^{2}-\beta ^{2})\frac{\left( \tilde{I}_{n}(z)\tilde{K}%
_{n}(z)\right) ^{\prime }}{z\tilde{I}_{n}^{\prime }(z)\tilde{K}%
_{n}^{\prime }(z)}\right] ,  \label{Egum}
\end{eqnarray}
for the total vacuum energy, and
\begin{equation}
\begin{array}{c}
\ds F=\frac{-2^{1-D}\pi ^{-D/2}}{a^{D}\Gamma (D/2-1)}\sum_{n=-\infty
}^{+\infty }\int_{ma}^{\infty }dz\, z\left(
z^{2}-m^{2}a^{2}\right) ^{D/2-2} \times
\\ \ds
\times
\left[ 2\beta
-4\xi +(z^{2}+n^{2}-\beta ^{2}+4\xi \beta )\frac{\left( \tilde{I}_{n}(z)%
\tilde{K}_{n}(z)\right) ^{\prime }}{z\tilde{I}_{n}^{\prime }(z)%
\tilde{K}_{n}^{\prime }(z)}\right]   \label{Fgum}
\end{array}
\end{equation}
for the total vacuum force acting per unit surface of the shell. In these
formulae we have introduced the notation
\begin{equation}
\tilde{f}(z)=z^{\beta }f(z),\qquad \beta =A/B  \label{nottilde}
\end{equation}
for a given function $f(z)$. The cases of Dirichlet and Neumann boundary
conditions are obtained taking the limits $\beta =\infty $ and $\beta =0$.
For the surface energies from (\ref{Esurfgum}) one obtains
\begin{equation}
\left\langle E^{{\rm (surf)}}\right\rangle _{SUB}^{{\rm (D)}}=
-\left\langle E^{%
{\rm (surf)}}\right\rangle _{SUB}^{{\rm (N)}}=\frac{\left( 1-4\xi
\right) L^{D-3}a^{2-D}}{2^{D-2}\pi ^{D/2-1}\Gamma
(D/2-1)}\sum_{n=-\infty }^{+\infty }\int_{ma}^{\infty
}dz\,z\left( z^{2}-m^{2}a^{2}\right) ^{D/2-2}. \label{EsurfDN}
\end{equation}
For the analytical continuation of the expression on the right we note that
\begin{equation}
\sum_{n=-\infty }^{+\infty }\int_{ma}^{\infty }dz\,z\left(
z^{2}-m^{2}a^{2}\right) ^{D/2-2}=\frac{1}{2}B(D/2-1,1-D/2)
(1+2\zeta _{R}(0))=0,  \label{sum1R}
\end{equation}
where $\zeta _{R}$ denotes the Riemann zeta function ($\ds\zeta
_{R}(s)=\sum_{n=1}^{\infty }n^{-s}$). As a result, we conclude that
{\em the surface energy is zero for Dirichlet and Neumann
scalars}. Due to the relation (\ref{sum1R}) we can omit $n$%
-independent terms in the subintegrands of (\ref{Esurfgum}),
(\ref{Fgum}) without changing the values of the sums. We can use
this to improve the convergence properties of the corresponding
$z$-integrals.

Let's present the total Casimir energy (\ref{Egum}) in another
equivalent form. For this, note that the functions $f(z)=
\tilde{I}_{n}(z)$, $\tilde{K}_{n}(z)$ satisfy the equation
\begin{equation}
z^{2}f^{\prime \prime }(z)+(1-2\beta )zf(z)-(z^{2}+n^{2}-\beta
^{2})f(z)=0. \label{eqItilde}
\end{equation}
Using this, it can be easily seen that
\begin{equation}
2\beta +(z^{2}+n^{2}-\beta ^{2})\frac{\left( \tilde{I}_{n}\tilde{K}%
_{n}\right) ^{\prime }}{z\tilde{I}_{n}^{\prime }\tilde{K}%
_{n}^{\prime }}=z\frac{\left( \tilde{I}_{n}^{\prime }\tilde{K}%
_{n}^{\prime }\right) ^{\prime }}{\tilde{I}_{n}^{\prime }\tilde{K}%
_{n}^{\prime }}+2(1-\beta )=z\frac{\left( \bar{I}_{n}\bar{K}%
_{n}\right) ^{\prime }}{\bar{I}_{n}\bar{K}_{n}} .
\label{transEtot}
\end{equation}
Thus, we obtain an equivalent representation for the total Casimir energy:
\begin{equation}
\left\langle E\right\rangle _{{\rm SUB}} =\frac{-L^{D-3}a^{2-D}}{2^{D-1}
\pi ^{D/2-1}\Gamma (D/2)}%
\sum_{n=-\infty }^{+\infty }\int_{ma}^{\infty
}dz(z^{2}-m^{2}a^{2})^{D/2-1} \, {\frac{d }{dz}} \ln\left(
\bar{I_n}(z) \bar{K_n}(z) \right) .  \label{Egum1}
\end{equation}
This expression is divergent in the given form. A method to extract
finite results will be explained below using the case of the $D=4$
massless scalar field as an example. The corresponding results for
arbitrary dimension and massive case will be presented elsewhere.

\section{Integrated Casimir energy per unit-length by mode sum}

Taking advantage of the usual prescription for evaluating the
eigenfrequency sum (see. e.g. ref.\cite{GR}), the integrated
Casimir energy per unit-length of an infinitely long cylinder is
expressed as
\begin{equation}
{\varepsilon }_{c}=\left. {\frac{1}{2}}\zeta _{\mbox{\scriptsize cyl}%
}(\sigma )\right| _{\sigma \rightarrow -1},
\end{equation}
where $\zeta _{\mbox{\scriptsize cyl}}(\sigma )$ is the zeta
function for the cylinder eigenfrequencies under the chosen
boundary conditions. Note that, in order to simplify the
procedure, we have not included the arbitrary mass scale
typically used in these problems (in fact, it will not be
necessary because the result will be finite). This zeta function
can be related to the one for a circle, with the same boundary
conditions and radius $a$, which we shall call $\zeta
_{\mbox{\scriptsize circ}}(s)$, and corresponds to the
restriction of the initial problem to a plane perpendicular to
the cylinder axis. Because of the cylindrical symmetry, the
eigenvalues of the wave operator ---including mass--- have the
form $(\gamma ^{2}+k^{2}+m^{2})^{1/2}$, where the $\gamma $'s are
the eigenfrequencies of
the restricted problem and $k\in \mbox{\bf R}$. Then, in a spacetime of $%
D=N+2+1$ dimensions,
\begin{equation}
\zeta _{\mbox{\scriptsize cyl}}(\sigma )=\int {\frac{d^{N}k}{(2\pi )^{N}}}%
\,\sum_{\gamma }\,\left( \gamma ^{2}+k^{2}+m^{2}\right) ^{-\sigma /2}={\frac{%
1}{(4\pi )^{N/2}}}{\frac{\Gamma \left( \frac{\sigma -N}{2}\right) }{\Gamma
\left( -{\frac{\sigma }{2}}\right) }}\zeta _{\mbox{\scriptsize circ}}(\sigma
-N)  \label{zcc0}
\end{equation}
with
\begin{equation}
\zeta _{\mbox{\scriptsize circ}}(s)\equiv \sum_{\gamma }(\gamma
^{2}+m^{2})^{-s/2}=\sum_{n}d_{n}\sum_{l}(\gamma_{n,l}^{2}+m^{2})^{-s/2},
\hspace{1cm}\gamma_{n,l}={\frac{\lambda _{n,l}}{a}}.  \label{zcc}
\end{equation}
In this section we are considering just one cylindrical surface, and these $%
\gamma$'s denote now the zeros of $\bar{J_n}(\gamma a)\bar{H}_n^{(1)}%
(\gamma a)$ (the first factor coming from the interior propagation and the
second from the exterior propagation). The $n$ index corresponds to angular
momentum, $d_{n}$ indicates each degeneracy, and the $l$ index labels the
different frequencies for a given $n$. Expanding (\ref{zcc}) around $\sigma
=-1$, one distinguishes two possibilities depending on $N$: \newline
\noindent 1. Even $N$ ($N=2p$)
\begin{equation}
\zeta _{\mbox{\scriptsize cyl}}(\sigma )={\frac{1}{2^{2p+1}\pi ^{p+1/2}}}%
\Gamma \left( -p-{\frac{1}{2}}\right) \zeta _{\mbox{\scriptsize circ}%
}(-2p-1).
\end{equation}
\noindent 2. Odd $N$ ($N=2p+1$)
\begin{equation}
\begin{array}{c}
\displaystyle\zeta _{\mbox{\scriptsize cyl}}(\sigma )={\frac{(-1)^{p}}{%
2^{2p}\pi ^{p+1}(p+1)!}} \\
\displaystyle\times \left[ {\frac{\zeta _{\mbox{\scriptsize circ}}(-2p-2)}{%
\sigma +1}}-{\frac{1}{2}}(\psi (-1/2)-\psi (p+2))\zeta _{%
\mbox{\scriptsize
circ}}(-2p-2)+{\zeta _{\mbox{\scriptsize circ}}}^{\prime }(-2p-2)+{\cal O}%
(\sigma +1)\right] .
\end{array}
\label{oddN}
\end{equation}
Later on we will focus on $D=3+1$ (i.e., $p=0$ case). Then, in situations
where $\zeta _{\mbox{\scriptsize circ}}(-2)=0$, (\ref{oddN}) reduces to
\begin{equation}
\zeta _{\mbox{\scriptsize cyl}}(s)={\frac{1}{2\pi }}{\zeta _{%
\mbox{\scriptsize circ}}}^{\prime }(-2).  \label{zeta-2van}
\end{equation}
Thus, one has to obtain the expansion of $\zeta _{\mbox{\scriptsize circ}%
}(s) $ up to linear terms in $(\sigma +1)=(s+2)$ (in general, $\sigma
+1=s+D-2$). By the Cauchy theorem, we may express the circle zeta function $%
\zeta _{\mbox{\scriptsize circ}}(s)$ as the following contour integrations
\begin{equation}
\begin{array}{lll}
\zeta _{\mbox{\scriptsize circ}}(s) & = & \displaystyle a^s \,
\sum_{n}d_{n}\, {\frac{1}{2\pi i}}\int_{C}dz\,(z^{2}+a^2 m^{2})^{-s/2}\,{\frac{d%
}{dz}}\ln \left( \bar{J_{n}}(z)\bar{H_n}^{(1)}(z)\right) \\
& = & \displaystyle a^s \, \sum_{n}d_{n}\,{\frac{s}{2\pi i}}%
\int_{C}dz\,(z^{2}+a^2  m^{2})^{-(s+2)/2}\,z\,\ln \left( \bar{J_{n}}(z)%
\bar{H_n}^{(1)}(z)\right) ,
\end{array}
\label{cauint}
\end{equation}
where $C$ is a circuit in the complex $z$-plane enclosing all the zeros of $%
\bar{J_{n}}(z)\bar{H_n}^{(1)}(z)$ (it is understood that in this
function we have now set $a=1$). In this way, we are already including the
contributions associated to internal and external propagation. Formulas (\ref
{cauint}) have the disadvantage of being valid for an $s$-domain in which $%
\mbox{\rm Re}\,s$ is positive, while we need to reach $s=-(D-2)$.
Just in order to study the formal relations between these
functions, we take the first form in (\ref{cauint}) and consider
a specific circuit. Let $C$ be the contour made of a vertical
line along the imaginary axis avoiding ---and leaving outside--- the cut of the$(z^{2}+a^2 m^{2})^{-s/2}$ function
(placed between $-iam$ and $+iam$),
and a semicircle of infinite radius to its right. After realizing which
parts yield no contribution, one finds
\begin{equation}
\zeta _{\mbox{\scriptsize circ}}(s)=\displaystyle a^{s} \, \sum_{n}\,d_{n}\,
{\frac{1}{\pi }}\sin \left( \frac{\pi s}{2}\right) \int_{am}^{\infty
}dx\,(x^{2}-a^{2}m^{2})^{-s/2}\,{\frac{d}{dx}}\ln \left( \bar{I_{n}}(x)%
\bar{K_{n}}(x)\right) .
\end{equation}
Now, using (\ref{zcc0}), one may formally write the integrated Casimir
energy per lateral unit-length as
\begin{equation}
{\frac{E_{C}}{L^{D-3}}}={\frac{1}{2}}\zeta _{\mbox{\scriptsize cyl}}(-1)=-{%
\frac{a^{2-D}}{2^{D-1}\pi ^{(D-2)/2}\Gamma \left( \frac{D}{2}\right) }}%
\,\sum_{n}\,d_{n}\,\int_{am}^{\infty }dx\,(x^{2}-a^{2}m^{2})^{-s/2}\,{\frac{d%
}{dx}}\ln \left( \bar{I_{n}}(x)\bar{K_{n}}(x)\right) .
\label{EcLform}
\end{equation}
Nevertheless, we have to stress that it is not valid to directly set $%
s=-(D-2)$ in order to obtain $E_{C}/L^{D-3}$. Analytic continuation to the
left of the present $s$-domain is required first. In the example below, we
explain a method for performing this type of process. Note, by the way, that
if one 'naively' sets $s=-(D-2)$ in (\ref{EcLform}), expression (\ref{Egum1}%
) is recovered.

\subsection{$m=0$, $D=3+1$ case}

In order to obtain the necessary analytic continuation to $s=-2$, we will
follow a different path, which has been explained e.g. in ref. \cite{LRap}.
Its application will be illustrated for the case $m=0$, $D=3+1$ (therefore $%
d_0=1$ and $d_n=2$ for $n\neq 0$). The mode sum is decomposed into sets of
partial waves with a defined angular momentum $n$, which, for convenience,
will be separately considered. Thus, we employ the notation
\begin{equation}
\zeta_{\mbox{\scriptsize circ}}(s)= \sum_{n} d_n \zeta_n(s)=
\sum_{n=-\infty}^{\infty} \zeta_n(s),
\end{equation}
where, taking the second form in (\ref{cauint}),
\begin{equation}
\zeta_n(s)=\sum_{l=1}^{\infty} \gamma_{n,l}^{-s} ={\frac{s }{2\pi i}}\int_C
dz \, z^{-s-1} \, \ln\left( \bar{J_n}(z) \bar{H_n}^{(1)}(z)
\right).
\end{equation}
A first step in the extension of the $s$-domain to the left is the study of
the asymptotic behaviour of the integrand for $|z|\to\infty$, given by $%
\displaystyle \bar{J_n}(z) \bar{H_n}^{(1)}(z) \sim -i z/2\equiv f_{%
\mbox{\scriptsize as}}(z)$. In view of this, we subtract from and add to the
integrand a term $z^{-s} \, \ln f_{\mbox{\scriptsize as}}(z)$. This move
leaves us with
\begin{equation}
\zeta_n(s)={\frac{s }{2\pi i}} \left[ \int_C dz \, z^{-s} \, \ln\left(\frac{%
\bar{J_n}(z) \bar{H_n}^{1}(z) }{f_{\mbox{\scriptsize as}}(z) }%
\right) +\int_C dz \, z^{-s} \, \ln f_{\mbox{\scriptsize as}}(z) \right] .
\end{equation}
Taking a contour $C$ as described
(now, since $m=0$, the cut line has reduced to the
point $z=0$)
we must conclude that the
second integral vanishes, because its integrand has no
singularity in the interior. As for the first integral, there are
nonvanishing contributions only from the vertical parts of $C$.
Parameterizing them adequately, we may write
\begin{equation}
\begin{array}{rcl}
\displaystyle\zeta_n(s) & = & \displaystyle a^s \, {\frac{s }{\pi}}\sin\left(%
\frac{ \pi s }{2 }\right) \, \int_{0}^{\infty} dx \, x^{-s} \, \ln L_n(x) ,
\\
L_n(x) & \equiv & \displaystyle -{\frac{2 }{x}} \left[ \bar{I_n}(x)
\bar{K_n}(x)\right] .
\end{array}
\end{equation}
This formula is slightly better than the initial one, as its domain of
validity has been shifted to the left: actually, it holds for $-1 < %
\mbox{\rm Re}\, s < 0$. However, it is not good enough, as we
have not arrived at $s=-2$ yet. To this end, we shall have to
perform even more subtractions. Before going on, we decompose
$\zeta_{\mbox{\scriptsize circ}}(s)$ into the contribution from
nonzero angular momenta ($n\neq 0$) and the contribution from
zero angular momentum ($n=0$):
\begin{equation}
\zeta_{\mbox{\scriptsize circ}}(s)= \zeta_{\mbox{\scriptsize circ}}^{%
{\scriptsize n\neq 0}}(s)+ \zeta_{\mbox{\scriptsize circ}}^{{\scriptsize n=0}%
}(s) .
\end{equation}
When $n\neq 0$ we may perform a rescaling $x \to nx$. Afterwards, summing
the internal and external contributions, we find
\begin{equation}
\zeta_n(s)= a^s \, n^{-s} \, {\frac{s }{\pi}} \sin\left(\frac{ \pi s }{2 }%
\right) \, \int_0^{\infty} dx \, x^{-s-1} \, \ln L_n(nx) .  \label{zmresc}
\end{equation}

We shall actually assume $B=1$ and take $A$ as the only variable.
This can be done whenever $B\neq 0$. Therefore, the only case
which cannot be described in this way is $B=0$, $A\neq 0$, i.e.,
purely Dirichlet conditions, which have already been considered
in ref.\cite{GR} (formula (34) in that paper). Studying the
uniform asymptotic expansions of the involved Bessel functions
(for $B\neq 0$) one realizes that
\begin{equation}
L_n(nx) \sim {\frac{ (1+x^2)^{1/2} }{x}} \left( B^2 +
\mbox{next-to-leading
terms} \right), \ \mbox{for large $nx$}.  \label{aslmx}
\end{equation}
This behaviour motivates (for $B=1$) the subtraction from and addition to
the integrand in eq. (\ref{zmresc}) of a term $\displaystyle x^{-s-1} \,
\ln\left(\frac{ (1+x^2)^{1/2} }{x }\right)$. Using then the result
\begin{equation}
\int_0^{\infty} dx \, x^{-s-1} \, \ln\left(\frac{ (1+x^2)^{1/2} }{x }\right)
={\frac{\pi }{2 s \sin\left(\frac{\pi s }{2 }\right) }} ,
\end{equation}
one obtains
\begin{equation}
\zeta_n(s)= a^s \left\{ n^{-s} \, {\frac{s }{\pi}} \sin\left(\frac{ \pi s }{%
2 }\right) \, \int_0^{\infty} dx \, x^{-s-1} \, \ln\left\vert {\frac{x }{%
(1+x^2)^{1/2} }} \, L_n(nx) \right\vert +{\frac{n^{-s} }{2}} \right\} .
\end{equation}
Hence,
\begin{equation}
\begin{array}{rcl}
\displaystyle\zeta_{\mbox{\scriptsize circ}}^{{\scriptsize n\neq 0}}(s) & =
& \displaystyle 2 \sum_{n=1}^{\infty} \zeta_n(s) \\
& = & \displaystyle a^s \left\{ 2 \, {\frac{s }{\pi}} \sin\left(\frac{ \pi s
}{2 }\right) \, \sum_{m=1}^{\infty} \, n^{-s} \, \int_0^{\infty} dx \,
x^{-s-1} \, \ln\left\vert {\frac{x }{(1+x^2)^{1/2} }} \, L_n(nx) \right\vert
\right. \\
&  & \displaystyle \hspace{1em}\left. +\zeta_R(s) \right\} .
\end{array}
\label{zmmneq00}
\end{equation}
All this comes from a formula valid for $\mbox{\rm Re}\, s$
between $-1$ and $0$. As already remarked, in order
to work at $s=-2$, further subtractions need to be done to the integral in (%
\ref{zmmneq00}). The result of applying such a technique is
\begin{equation}
\begin{array}{lll}
\displaystyle \zeta_{\mbox{\scriptsize circ}}^{{\scriptsize n\neq 0}}(s) & %
\displaystyle =a^s & \displaystyle \left\{ {\frac{s}{\pi}}\sin\left(\frac{%
\pi s }{2 }\right)\left[ 2\sum_{n=1}^{\infty} n^{-s} \, {\cal S}_n(N,s)
+2\sum_{j=1}^{N/2} \zeta_R(s+2j) \, \sum_k {\cal U}_{2j,k} \, \mbox{B}\left(
{\frac{s+k}{2}}, -{\frac{s}{2}} \right) \right] \right. \\
&  & \displaystyle \left. +\zeta_R(s) \right\},
\end{array}
\label{zcircnneq0}
\end{equation}
where
\begin{equation}
{\cal S}_n(N,s)=\int_0^{\infty} dx \, x^{-s} \left[ \ln \left\vert {\frac{x}{%
(1+x^2)^{1/2}}} \, L_n(nx) \right\vert -2\sum_{j=1}^{N/2} {\frac{{\cal U}%
_{2j}(t(x)) }{n^{2j} }} \right] .
\end{equation}
In formula (\ref{zcircnneq0}), the objects called ${\cal
U}_{2j}(t)$ are polynomials in $t$, where
\begin{equation}
t(x)={\frac{1 }{(1+x^2)^{1/2} }},
\end{equation}
which are constructed step by step when obtaining the asymptotic
approximation
\begin{equation}
\ln \left\vert {\frac{x}{(1+x^2)^{1/2}}} \, L_n(nx) \right\vert \sim
2\sum_{j=1}^{N/2} {\frac{{\cal U}_{2j}(t(x)) }{n^{2j} }},
\end{equation}
for large $nx$ ---in fact, this is like a further elaboration of (\ref{aslmx}%
)---. Then, the ${\cal U}_{2j,k}$'s are the present coefficients of these
polynomials, in other words,
\begin{equation}
{\cal U}_{2j}(t)= \sum_k {\cal U}_{2j,k} \, t^k,
\end{equation}
where the $k$-index range corresponds to the existing
coefficients. The whole process for obtaining this type of
approximation has already been described in detail in
refs.\cite{Rprd95,LRap} (see also \cite{AS}). By comparison, one
may reason that these ${\cal U}_{2j}(t)$'s have to be the same as
the
even-$j$ ${\cal U}_j^{I, {\cal R}}(t)$'s in formula (3.37) of ref. \cite{LRap}%
, after the replacement $\alpha\to A$. The $N$ variable denotes the order of
the subtractions performed in order to remove negative-$s$ poles from the
initial integral. If $N$ is large enough ---and in our case this means just $%
N\ge 2$--- the ${\cal S}_n(N,s)$ integral is finite at $s=-2$. When $N$ is
further increased, the value of the numerical integration ${\cal S}_n(N,s)$
gets smaller and smaller while the algebraic $n$-sum (easier to evaluate)
grows accordingly. In practice, we take $N=$8 or 10.

The $n=0$ contribution has to be dealt with separately, as a step in the
process for $n\neq 0$ involved rescaling by $n$, and cannot be applied now.
The same philosophy as in refs. \cite{LRap,GR} will now be adopted. We will
just perform the necessary subtraction to isolate the $s=-2$ divergence from
the integral, leaving the rest to numerical evaluation. The result is
expressed in the form
\begin{equation}
\zeta_{\mbox{\scriptsize circ}}^{{\scriptsize n=0}}(s)= a^s \, {\frac{s}{\pi}%
}\sin\left(\frac{\pi s }{2 }\right)\left[ {\cal R}_0(s)-\left( A^2-A+3/8
\right){\frac{1 }{2}} \mbox{B}\left( {\frac{s+2}{2}}, -{\frac{s}{2}} \right) %
\right],  \label{zcircneq0}
\end{equation}
where the integral
\begin{equation}
{\cal R}_0(s)=\int_0^{\infty} dx \, x^{-s} \left[ \ln \left\vert L_0(x)
\right\vert +\left( A^2-A+3/8 \right) \, t^2(x) \right]
\end{equation}
is, by construction, finite at $s=-2$. Therefore, ${\cal R}_0(-2)$ can be
calculated numerically for given values of $A$ ($B=1$).

Thus, the circle zeta function is the sum of (\ref{zcircnneq0})
and (\ref {zcircneq0}). At $s=-2$, (\ref{zcircnneq0}) has a
nonzero finite value coming from the pole of $\mbox{B}\left(
{\frac{s+2}{2}}, -{\frac{s}{2}} \right)$ multiplied by the zero
of ${\frac{s}{\pi}}\sin\left(\frac{\pi s }{2
}\right)$ at $s=-2$. This nonzero value depends also on the involved ${\cal U%
}_{2j,2}$ coefficient(s) and on $\zeta_R(s+2j)$. Now, making a
similar reasoning for (\ref{zcircneq0}), we realize that the
$s=-2$ value of this expression is just $-{\frac{1
}{a^2}}(A^2-A+3/8)$. In fact, algebraic calculation shows that
the finite value of (\ref{zcircnneq0}) at $s=-2$
amounts to the same quantity with opposite sign. Therefore, $\zeta_{%
\mbox{\scriptsize circ}}(-2)=0$, and one can safely apply formula (\ref
{zeta-2van}). In these circumstances, the matter reduces to finding ${\zeta_{%
\mbox{\scriptsize circ}}}^{\prime}(-2)$ by expanding the sum of (\ref
{zcircnneq0}) and (\ref{zcircneq0}) around $s=-2$. Once this has been done,
the integrated Casimir energy per unit-length is simply
\begin{equation}
{\varepsilon}_c= {\frac{1 }{4\pi}} {\zeta_{\mbox{\scriptsize circ}}}%
^{\prime}(-2).
\end{equation}
Results for a given $A$-range around $A=0$ (purely Neumann) are shown in
Fig. \ref{figecA} below, where we have plotted the numerical values of ${\varepsilon}_c
\cdot a^2$ as a function of $A$.
\begin{figure}[htbp]
\begin{center}
\begin{tabular}{c}
\psfig{figure=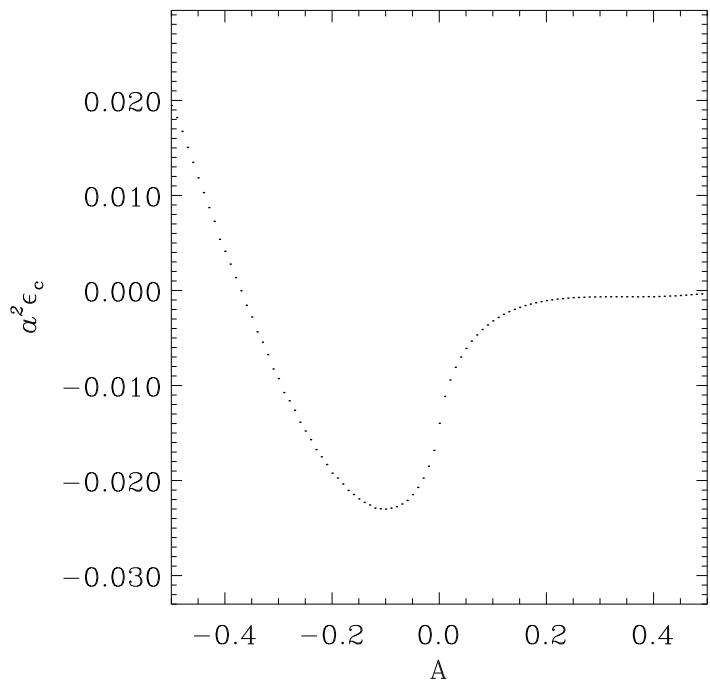,width=10cm,height=10cm}
\end{tabular}
\end{center}
\caption{Total integrated Casimir energy per unit-length, multiplied by $a^2$%
, for $B=1$ as a function of the $A$ coefficient, in a region
around $A=0$. Note the presence of a minimum near $A=-0.010$. The
value for $A=0$ (purely Neumann b.c.) is of $-0.014176...$ in
agreement with formula (44) of ref.{\ \protect\cite{GR}}. }
\label{figecA}
\end{figure}

\section{Final remarks}

In the present paper, the vacuum ---or zero-point--- energy of scalar fields
obeying Robin boundary conditions on cylindrical surfaces has been studied.
Local quantities have been evaluated starting from vacuum expectation values
of energy momentum tensors and applying generalized Abel-Plana summation
methods supplemented with a suitable cutoff function. Integrated energies
seem to be more easily found from the eigenfrequency sum which determines
the Casimir effect. In that case, zeta-function regularization has proven to
be an adequate tool. However, both approaches constitute different aspects
of one underlying idea. In fact, we have shown that the two methods lead to
identical formal expressions ---formulas (\ref{Egum1}) and (\ref{EcLform})
at $s=-(D-2)$.

We have obtained an expression for the integrated energy inside
the cylinder. In general, the total vacuum energy does not
coincide with the integral of the volumic density, because there
is a purely superficial contribution, located at the boundary
itself, which has to be added to the volumic part. Moreover, the
contributions associated to the modes propagating in the exterior
of the cylinder has also been considered. In both cases, there is
a part coming from the surface ('viewed' from inside, in the
first, and from outside, in the second). The total Casimir energy
arises as the sum of all these contributions, and may also be
obtained by eigenfrequency summation. As an example, we have
chosen the massless case in $D=3+1$ dimensions for application of
the zeta function method. The result for a given parameter range
is shown in fig. \ref{figecA}, which displays the presence of a
local minimum near (but not coinciding with) the point where
Robin conditions become purely Neumann conditions. This indicates
that, in a theoretical space of possible boundary conditions,
some linear combinations are energetically privileged.


\begin{thebibliography}{99}
\bibitem{PMG}  G. Plunien, B. M{\"{u}}ller and W. Greiner, {\it Phys. Rep. }
{\bf 134}, 87 (1986); V.M. Mostepanenko and N.N. Trunov, {\it The Casimir
Effect and its Applications}, Oxford Univ. Press, 1997; K.A. Milton, {\it %
The Casimir Effect: Physical Manifestations of the Zero-Point Energy},
hep-th/9901011; {\it Dimensional and Dynamical Aspects of the Casimir
Effect: Understanding the Reality and Significance of Vaccum Energy},
hep-th/0009173.

\bibitem{Ambjorn2} J. Ambj{\o}rn and S. Wolfram, {\it Ann. Phys.}
{\bf 147} (1983) 33.

\bibitem{Moss} I. G. Moss, {\it Class. Quantum Grav.} {\bf 6}
(1989) 759.

\bibitem{Espos} G. Esposito, A. Yu. Kamenshchik and G. Polifrone,
{\it Euclidean Quantum Gravity on Manifolds with Boundariy},
Kluwer, 1997.

\bibitem{Fishbane} P. Fishbane, S. Gaziorovich and P. Kauss,
{\it Phys. Rev.} {\bf 36} (1987) 251; {\bf 37} (1988) 2623.

\bibitem{Barbash} B. M. Barbashov and V. V. Nesterenko,
{\it Introduction to the Relativistic String Theory}, World
Scientific, Singapore, 1990.

\bibitem{Ambjorn1} J. Ambj{\o}rn and S. Wolfram, {\it Ann. Phys.}
{\bf 147} (1983) 1.

\bibitem{DRM}  L.L. De Raad Jr. and K.A. Milton, {\it Ann. Phys.} {\bf 136}
(1981) 229.

\bibitem{GR}  P. Gosdzinsky and A. Romeo, {\it Phys. Lett. } {\bf B 441}
(1998) 265.

\bibitem{MNN}  K.A. Milton, A.V. Nesterenko and V.V. Nesterenko, {\it Phys.
Rev. } {\bf D 59} (1999) 105009.

\bibitem{Sah1cyl}  A. A. Saharian, {\it Izv. AN Arm. SSR. Fizika} {\bf 23}
(1988) 130 ( {\it Sov. J. Contemp. Phys.} {\bf 23} (1988) 14).

\bibitem{Sah2cyl}  A. A. Saharian, {\it Dokladi AN Arm. SSR} {\bf 86} (1988)
112 ({\it Reports NAS RA}, in Russian).

\bibitem{Sahrev}  A. A. Saharian, {\it Izv. AN Arm. SSR. Matematika} {\bf 22}
(1987) 166 ({\it Sov. J. Contemp. Math.} {\it Analysis} {\bf 22} (1987) 70);
A. A. Saharian, The generalized Abel-Plana formula. Applications to Bessel
functions and Casimir effect. Preprint IC/2000/14, hep-th/0002239.

\bibitem{Deutsch}  D. Deutsch and P. Candelas, {\it Phys. Rev.} {\bf D20}
(1979) 3063.

\bibitem{RomSah}  A. Romeo and A. A. Saharian, Casimir effect for scalar
fields under Robin boundary conditions on plates, Preprint hep-th/0007242.

\bibitem{Sahsph} A. A. Saharian, Scalar Casimir effect for
D-dimensional spherically symmetric Robin boundaries,
hep-th/0012185.

\bibitem{Birrel} N. D. Birrel and P. C. W. Davies,
{\it Quantum fields in curved space}, Cambridge University Press,
1982.

\bibitem{KCD}  G. Kennedy, R. Critchley and J.S. Dowker, {\it Ann. Phys. }
{\bf 125} (1980) 346.

\bibitem{PrudnikovSF}  A. P. Prudnikov, Yu. A. Brychkov, O. I. Marichev,
{\it Integrals and series. v.2. Special functions.} 1986.

\bibitem{Rprd95}  A. Romeo, {\it Phys. Rev.} {\bf D 52} (1995) 7308-7314.

\bibitem{LRap}  S. Leseduarte and A. Romeo, {\it Ann. Phys.} {\bf 250}
(1996) 448.

\bibitem{AS}  M. Abramowitz and I. A. Stegun, {\it Handbook of Mathematical
functions}, National Bureau of Standards, Washington D.C., 1964.

\end{thebibliography}
\end{document}